\documentclass[preprint,aps]{revtex4}

\usepackage{graphicx}
\usepackage{dcolumn}
\usepackage{bm}

\begin{document}

\topmargin=0.1in

\preprint{}

\title{Structure and Dynamics of Charmonium}

\author{Kevin L. Haglin}
\homepage{http://feynman.stcloudstate.edu/haglin}
\affiliation{Department of Physics, Astronomy and Engineering
Science\\St. Cloud State University, St. Cloud, MN 56301, USA}

\date{\today}

\begin{abstract}
We apply a confining q$\bar{\mbox{q}}$ potential to charmonium
and open charm states in order to model wave functions and
to begin studying structure.   Results (in momentum space) provide
form-factor input to a four-flavor 
effective chiral Lagrangian which models dynamics of charmonium in hot 
hadronic
matter.  Estimates are made for $J$/$\psi$ dissociation cross sections 
and rates within a fireball.   Our study attempts to improve on 
previous comover suppression models since it includes gauge-invariant 
form-factor formalism constrained by quark-model phenomenology.
\end{abstract}

\pacs{PACS: 12.39.Fe, 12.39.Hg, 13.75.Lb, 14.40Lb}

\maketitle

\section{Introduction}\label{intro}
Signals of momentary quark and gluon deconfinement in high-energy
heavy-ion reactions are studied today more aggressively than ever
before owing to current experimental activities at the Relativistic 
Heavy-Ion Collider (RHIC)\cite{Nassau_talks}.   Hard probes represent 
a particular 
piece to the overall puzzle whose goal upon assembly is to 
fully understand the
strongly interacting many-body dynamics of ultra-relativistic
heavy-ion collisions, complete with definite evidence for
quark-gluon plasma (QGP) formation and propagation.   The hard probes
provide complementary information to 
such softer probes as photon and dilepton production.  
Very briefly for charm, the  idea of Matsui 
and Satz\cite{ms86}
proposes that in events where QGP is formed, screening effects
tend to break apart the $c\bar{c}$ states leaving a
``gap'' between observed charmonium and expected.  It is 
very convenient to look for evidence of charmonium breakup by
studying the mass distribution of muon pairs and trimming 
away the background of non-$J$/$\psi$ contributions.
There have already been suggestions that such a comparison might
suggest glimpses of QGP\cite{na50}.  

Meanwhile, several authors have begun to systematically assess comover
absorption using a variety of different approaches in order to improve
understanding of the background due to such
effects as light meson plus charmonium interactions leading to
breakup\cite{mueller,haglin,khcg01,oh,lin,wong,blaschke}.  Breakup of 
this type could possibly be misinterpreted as subhadronic effects
when indeed, it is ordinary hadronic many-body physics.  The aim
of this study is to further refine estimates of hadronic cross sections for
breakup of $J$/$\psi$ by constraining form factors with quark-model 
phenomenology and confining potentials,
and then to use the form factors
in a four-flavor chiral Lagrangian.

Our article is organized in the following way.  We discuss in 
Sect.~\ref{quarkmodel} the confining potential and resulting
meson wave functions.  They provide information on the hadronic
form factors to be later used in charmonium dynamics.  Sect.~\ref{chirall}
includes a brief summary of the four-flavor chiral Lagrangian used
to model the dynamics of the hadronic matter constituents.
It also discusses gauge-invariant implementation
of finite hadron size effects, namely form factors.  Results
are presented and discussed in Sect.~\ref{results} and 
finally, Sect.~\ref{concl} summarizes and concludes.

\section{Confining Potential}\label{quarkmodel}  
Accomplishments of quark-model studies include rather detailed
comparisons of calculated hadron spectra versus observed for a long list 
of light mesons.
The models are constrained as firmly as possible by confronting
such spectroscopic details as masses and decay widths.  We take 
a recent result from the literature\cite{roberts} and extend it beyond 
the light mesons to include $D$'s $\eta_{c}$, $J$/$\psi$ and 
$\psi^{\,\prime}$ (no isospin dependence is included, so we do not
mention $\chi_{c}$ states explicitly).  The potential is
\begin{equation}
V_{ij} = -{\kappa\over r}+\lambda\,r^{p}+\Lambda+{2\pi\kappa^{\,\prime}
\over\,3m_{i}m_{j}}{\mbox{exp}(-r^{2}/r_{0}^{2})\over\,\pi^{3/2}r_{0}^{3}}\,
\vec{\sigma}_{i}\cdot\vec{\sigma}_{j},
\label{vij}
\end{equation}
where the range $r_{0}$ of the hyperfine term is taken to
be mass dependent 
\begin{equation}
r_{0}(m_{i},m_{j}) = A\left({2m_{i}m_{j}\over\,m_{i}+m_{j}}\right)^{-B}.
\label{rnot}
\end{equation}
A set of parameters is chosen to give the usual Coulomb plus linear
form for the central part
\begin{eqnarray}
m_{u} = m_{d} = 0.315 \mbox{\ GeV}; \quad m_{s} = 0.577 
\mbox{\ GeV};& \ &  
m_{c} = 1.836 \mbox{\ GeV}
\nonumber\\
\kappa = 0.5069; \quad
\kappa^{\,\prime} = 1.8609; & \ & \lambda = 0.1653 
\mbox{\ GeV}^{2}; \quad p = 1
\nonumber\\
\Lambda = -0.8321 \mbox{\ GeV}; \quad B = 0.2204; & \ &   
A = 1.6553 \mbox{\ GeV}^{B-1}.
\end{eqnarray}

By numerically solving the Schr\"odinger equation bound state wave
functions are obtained,
from which meson masses and rms radii are readily 
computed.  Results for a selected set of 
mesons are listed in table~\ref{tab1}.  Much of the motivation to
do subnuclear structure in this way is to use the information as
form-factor input to an effective Lagrangian description of the
strongly-interacting hadrons.   Typically, one views the
spatial density as the Fourier transform of a (momentum space)
distribution---the form factor.  As is usually assumed in 
field theoretic modeling of this type, and is consistent with quark counting
rules, a monopole structure is used.  We extrapolate from massless
to massive probes, and use a monopole-charge-form-factor-inspired expression
(only meant as an indicator rather than a consistent model calculation)
for the cutoff or off-shellness parameter
\begin{equation}
\Lambda = \sqrt{m^{2} + {6\over \langle\,r^{2}\rangle}}.
\end{equation}
The monopole form factor to be discussed later uses $\Lambda$ in 
attempts to describe three-point vertices where a meson of mass $m$ and rms
radius $\sqrt{\langle\,r^{2}\,\rangle}$ is forced to go off shell.  In 
rough terms, the size of the interaction vertex is inversely proportional
to the cutoff parameter.  And again, it is the off shell particle
in a three point vertex which governs the physics.

\begin{table}[hb] 
\vspace*{-12pt}
\caption[]{Masses, rms radii and
form-factor cutoff parameter for a select set of 
mesons.}\label{tab1}
\vspace*{-14pt}
\begin{center}
\begin{tabular}{ccccc}
\hline\\[-10pt]
Meson & Mass (MeV) [Obs.] & Mass (MeV) [Calc.] & 
$\sqrt{\langle\,r^{\,2}\,\rangle}$ (fm) & $\Lambda$ (GeV) \\ 
\hline\\[-10pt]
$\pi$ & 138 & 138 & 0.59 & 0.80 \\
$K$ & 496 & 490 & 0.59 & 0.96 \\
$\rho$ & 769 & 770 & 0.92 & 0.93 \\
$K^{*}$ & 894 & 904 & 0.81 & 1.07 \\
$\phi$ & 1019 & 1021 & 0.70 & 1.23 \\
$a_{1}$ & 1230 & 1208 & 1.24 & 1.29 \\
$D$ & 1867 & 1862 & 0.61 & 2.03 \\
$D^{*}$ & 2008 & 2016 & 0.70 & 2.12 \\
$\eta_{c}$ & 2980 & 3005 & 0.37 & 3.25 \\
$J$/$\psi$ & 3097 & 3101 & 0.40 & 3.32 \\
$\psi^{\,\prime}$ & 3686 & 3641 & 0.79 & 3.73 \\
\hline 
\end{tabular}
\end{center}
\end{table}

\section{Four-Flavor Chiral Lagrangian}\label{chirall}  

In the absence of firm experimental constraints on a full set of
mesonic interactions involving strangeness and charm, effective
theories, with a specified chirally symmetric interaction, are quite
useful.  Indeed, there has been a renewed interest in these
approaches since full
understanding and control of the nonperturbative effects confinement
necessitates is still beyond grasp.  Chiral symmetry and current
conservation represent minimum requirements for any effective hadronic
field theory.  We take such
an approach here by extending the usual two-flavor chiral Lagrangian
to not three, but a four-flavor set of fields.  The strange quark
mass being greater than up and down quark masses probably already brings 
about some limitations for the Lagrangian's usefulness, and the charmed
quark mass certainly breaks the symmetry to a deeper extent.
And yet, it is not unreasonable to relegate this breaking to the mass
terms, and insist that the interaction remain fully symmetric. 

The full details, starting from the nonlinear sigma
model, introducing vector and axial vector fields into gauge covariant 
derivatives and then subsequently gauging away all of the axial
degrees of freedom, have been published elsewhere\cite{khcg01}.  We therefore
include here the interaction terms alone.  They are written very compactly
as 
\begin{eqnarray}
{\cal L\/}_{int}^{I} & = & ig\,{\rm Tr}
\left(\rho_{\mu}\left[\partial^{\mu}\phi,\phi\right]\right)
- {g^{2}\over 2}{\rm Tr}(\left[\phi,\rho^{\mu}\right]^{2})
+ ig\,{\rm Tr}\left(\partial_{\mu}\rho_{\nu}\left[\rho^{\mu},\rho^{\nu}
\right]\right)
+ {g^{2}\over 4}{\rm Tr}(\left[\rho^{\mu},\rho^{\nu}\right]^{2})\quad
\label{ellint}
\end{eqnarray}
where $\phi$ and $\rho^{\mu}$ are 4$\times$4 matrices with entries 
containing pseudoscalar and vector fields, respectively.

Since a strict chiral symmetry is respected, there remains just
one (chiral) coupling constant to fix.  We do so by making certain the
rho meson is correctly described.   The choice $g$ = 4.3 gives 
$\Gamma_{\rho}$ = 151 MeV at the pole mass of 770 MeV, and gives 
decay widths for strangeness and charm excitations listed in 
Table~\ref{tab2}.

\begin{table}[hb] 
\vspace*{-12pt}
\caption[]{Model prediction for the $K^{*}$ 
and $D^{*}$ decay widths as compared to experiment.
}\label{tab2}
\vspace*{-14pt}
\begin{center}
\begin{tabular}{ccc}
\hline
particle
& \ Chiral Lagrangian\
 & Experiment \\
\hline
K(892)$^{0}$ & 44.5 MeV & 50.5 $\pm$ 0.6 MeV \\ 
K(892)$^{\pm}$ & 44.5 MeV & 49.8 $\pm$ 0.8 MeV \\ 
D(2007)$^{0}$ & 10.1 keV & $<$ 2.1 MeV, 90\% CL\\ 
D(2010)$^{\pm}$ & 21.1 keV & 96 $\pm$ 4 $\pm$ 22 keV\protect{\cite{cleo}} \\
\hline
\end{tabular}
\end{center}
\end{table}

The very recent charged $D^{*}$ decay measurement coming in at 96 keV
allows for a $D^{*}D\pi$ coupling constant to be fully twice as large
as the chiral symmetry proposes.  This would increase the dissociation
cross section be precisely a factor of two.  For now, however, we
adhere to the chiral symmetry prediction.

The interactions identified in Eq.~(\ref{ellint}) do not include
anomalous processes of type vector-vector-pseudoscalar.
We therefore extend the set of interactions by introducing
\begin{equation}
{\cal L}^{II}_{int} = g_{VV\phi}\,{\rm Tr}\left(
\epsilon_{\mu\nu\alpha\beta}\partial^{\mu}V^{\nu}\partial^{\alpha}
V^{\beta}\phi\right),
\end{equation}
with coupling constants constrained individually via vector meson dominance.  
One of the channels now open with
${\cal L\/}^{II}$ is $J/\psi + \pi\rightarrow \eta_{c} + \rho$, an
important contributor.

\subsection{Form Factors}\label{fff}

Effective theories attempt to model composite
objects which necessarily have finite extent, and are therefore
responsible for finite-sized interaction vertices.  Three-point 
functions, representing full details of the 
interactions including loop effects to arbitrary order, 
are notoriously difficult to handle consistently.   One-boson-exchange
models, which involve three- and possibly four-point vertices, approximate
the full calculation.  Since in such models at most one particle
per vertex goes off shell, a Lorentz invariant form factor accounts
for dressing the vertex.  So, rather than attempt to calculate higher-order
contributions, we assume a specific
monopole form for the momentum dependent vertex coupling
``constants'' , and use the off-shell or cutoff parameters from 
Table~\ref{tab1}.  All three-point vertices are therefore given the
following monopole
\begin{equation}
h(t) = {\Lambda^{2}\over\,\Lambda^{2}\,+\,|\,t\,-\,m^{2}\,|}\,,
\end{equation}
where $t$ here is the squared four-momentum of the off-shell particle.
Notice that when $t\rightarrow m^{2}$, then $h\rightarrow 1$, which
is indeed how the pole coupling constants are all defined.

Four-point functions are modified from their typical $g^{\mu\nu}$
form to the most general linear combination of all possible 
lowest-order Lorentz invariant
structures constructible out of the external momenta.  Specifically,
in the reaction $J/\psi+\pi\rightarrow\,\overline{D}+D^{*}$, the 
four-point vertex becomes
\begin{eqnarray}
\Gamma^{\mu\nu} & = & A\,g^{\mu\nu} 
+ B\,\left(p_{\overline{D}}^{\mu}\,p_{\pi}^{\nu} + 
p_{\pi}^{\mu}\,p_{\overline{D}}^{\nu}\right)
+ C\,\left(p_{D^{*}}^{\mu}\,p_{\pi}^{\nu} + 
p_{\pi}^{\mu}\,p_{D^{*}}^{\nu}\right)
+ D\,\left(p_{\pi}^{\mu}\,p_{\pi}^{\nu} + 
p_{\overline{D}}^{\mu}\,p_{\overline{D}}^{\nu}\right)
\nonumber\\
&\  & \quad + E\,\left(p_{\pi}^{\mu}\,p_{\pi}^{\nu} + 
p_{D^{*}}^{\mu}\,p_{D^{*}}^{\nu}\right),
\end{eqnarray}
and then the expansion coefficients $A$--$E$ are chosen so that the 
overall scattering
amplitude is fully gauge invariant, $\partial_{\mu}{\cal M\/}^{\mu}
= 0$.  The choice is not unique; but on the other hand, gauge invariance
alone is not enough to uniquely fix an amplitude.  It does however
represent an absolute minimum requirement of {\it any\/} reliable
model. 
 
\section{Results}\label{results}  

With all the interactions identified in the model, all the
vertices constrained as much as possible, a list of reactions involving
light meson plus $J/\psi$ can be enumerated and calculated.  We begin 
looking at
$\pi$, $K$, $\eta$, $\rho$, $\omega$, $K^{*}$, and so on. 

\subsection{Cross Sections}\label{xsects}
\begin{figure}
\includegraphics{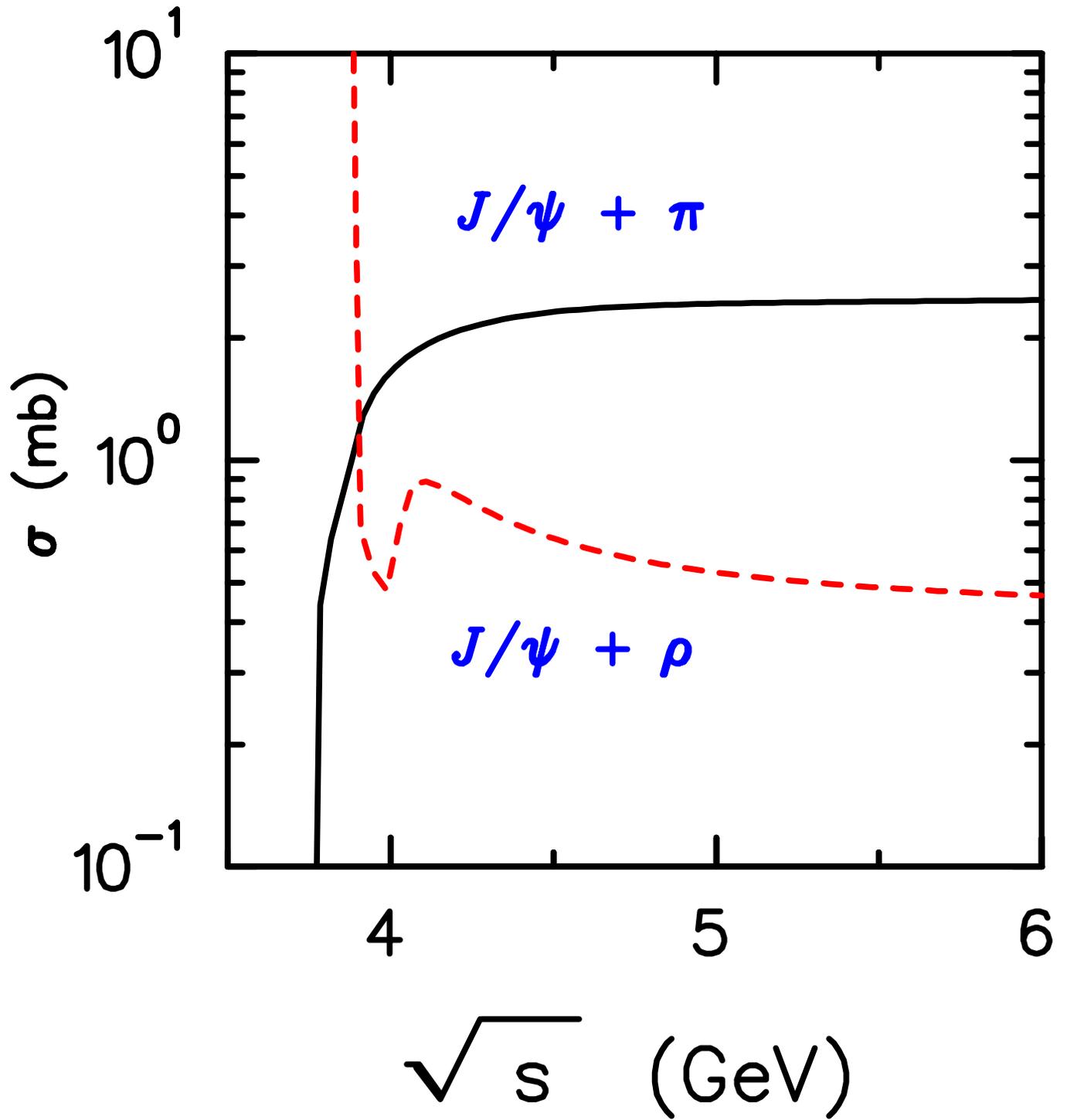}
\caption{\label{fig1}Cross sections for pion- and rho-induced 
dissociation. For complete details on all final states 
see Ref.~\protect{\cite{khcg01}}.}
\end{figure}

The required dynamical quantity for estimates of $J/\psi$ production
and possible suppression
are the breakup cross sections for the
individual reactions.  There are too many specific channels
studied here to completely discuss them all.  Instead, we show 
the two leading contributors in Fig.~\ref{fig1}.   The pion-induced 
reactions involve $D+\overline{D}^{*}$,
$\overline{D}+{D}^{*}$, and $\eta_{c}+\rho$ final states and the
rho-induced reactions involve $D+\overline{D}$, $D^{*}+\overline{D}^{*}$
and $\eta_{c}+\pi$ final states.   In a rough summary, the cross 
sections range from a few tenths to a few millibarns.

\subsection{Dissociation Rates}\label{rates}

\begin{figure}
\includegraphics{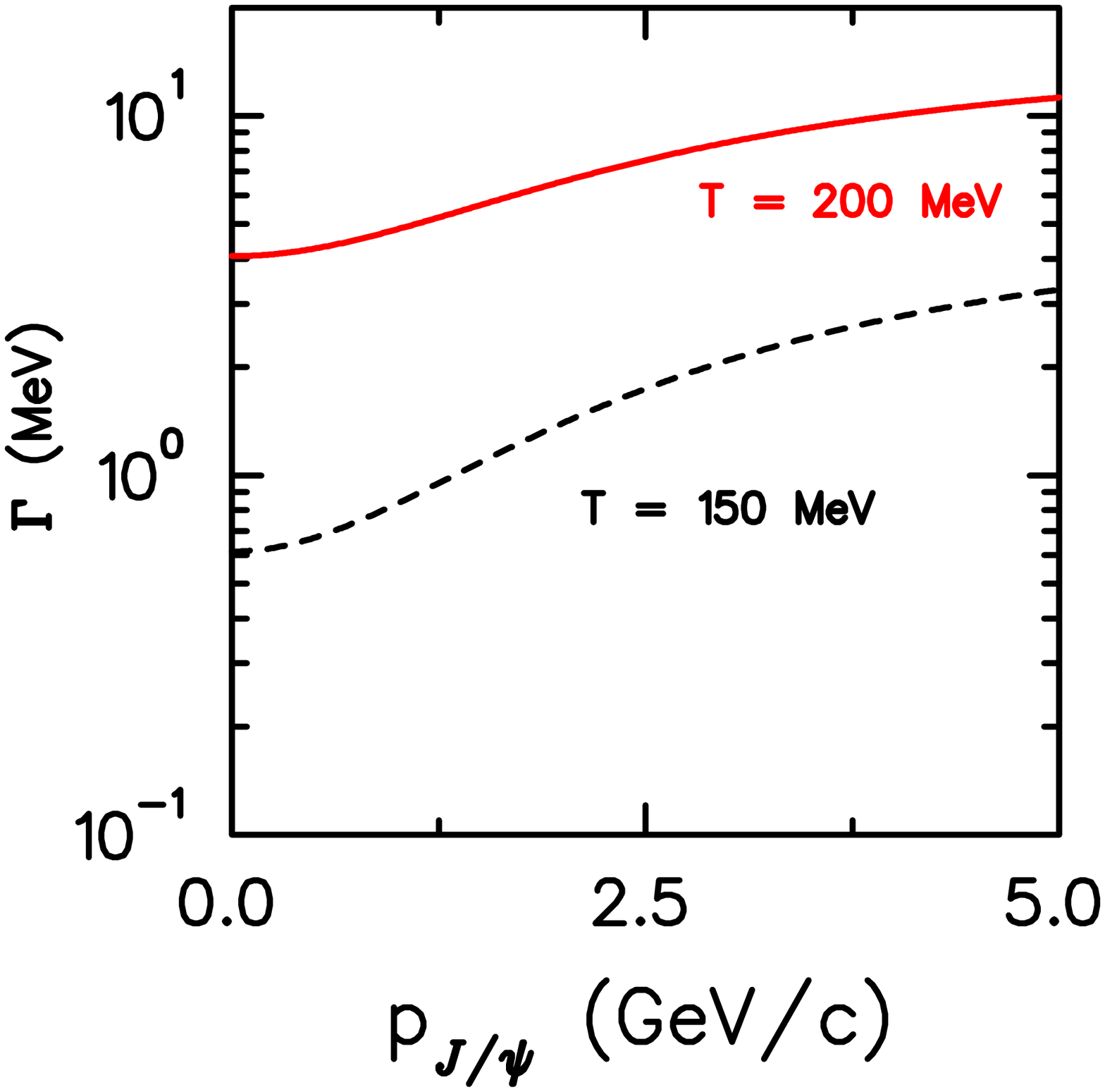}
\caption{\label{fig2}Dissociation rates in a fireball at fixed temperatures,
150 and 200 MeV as a function of $J/\psi$ (nonthermal) momentum.  Rates 
included $\pi$, $K$, $\rho$ and $K^{*}$-induced
breakup.}
\end{figure}

Consider the reaction $J/\psi + b \rightarrow 1 + 2$.  The rate
within a fireball for this to occur is the following.

\begin{eqnarray}
d\/\Gamma_{J/\psi} &=& {1\over\,2E_{J/\psi}}\,d_{b}{d^{3}p_{b}\over
(2\pi)^{3}2E_{b}}\, f_{b}\,|\overline{\cal\,M}|^{2}\,\tilde{f}_{1}
\,\tilde{f}_{2}(2\pi)^{4}\delta^{4}(p_{J/\psi} + p_{b} - p_{1}
- p_{2})
\label{dgamma}\\
\nonumber 
& \ & \quad\quad\times{d^{3}p_{1}\over(2\pi)^{3}2E_{1}}
{d^{3}p_{2}\over(2\pi)^{3}2E_{2}},
\end{eqnarray}
where $d_{b}$ is the degeneracy of species $b$, $f_{b}$ is the
Bose-Einstein distribution, and $\tilde{f} \equiv 1+f$ to account
for the medium.  We show in Fig.~\ref{fig2} the total dissociation
rates at two fixed temperatures as functions of the $J/\psi$ momentum.

\subsection{Flow}\label{frates}

\begin{figure}
\includegraphics{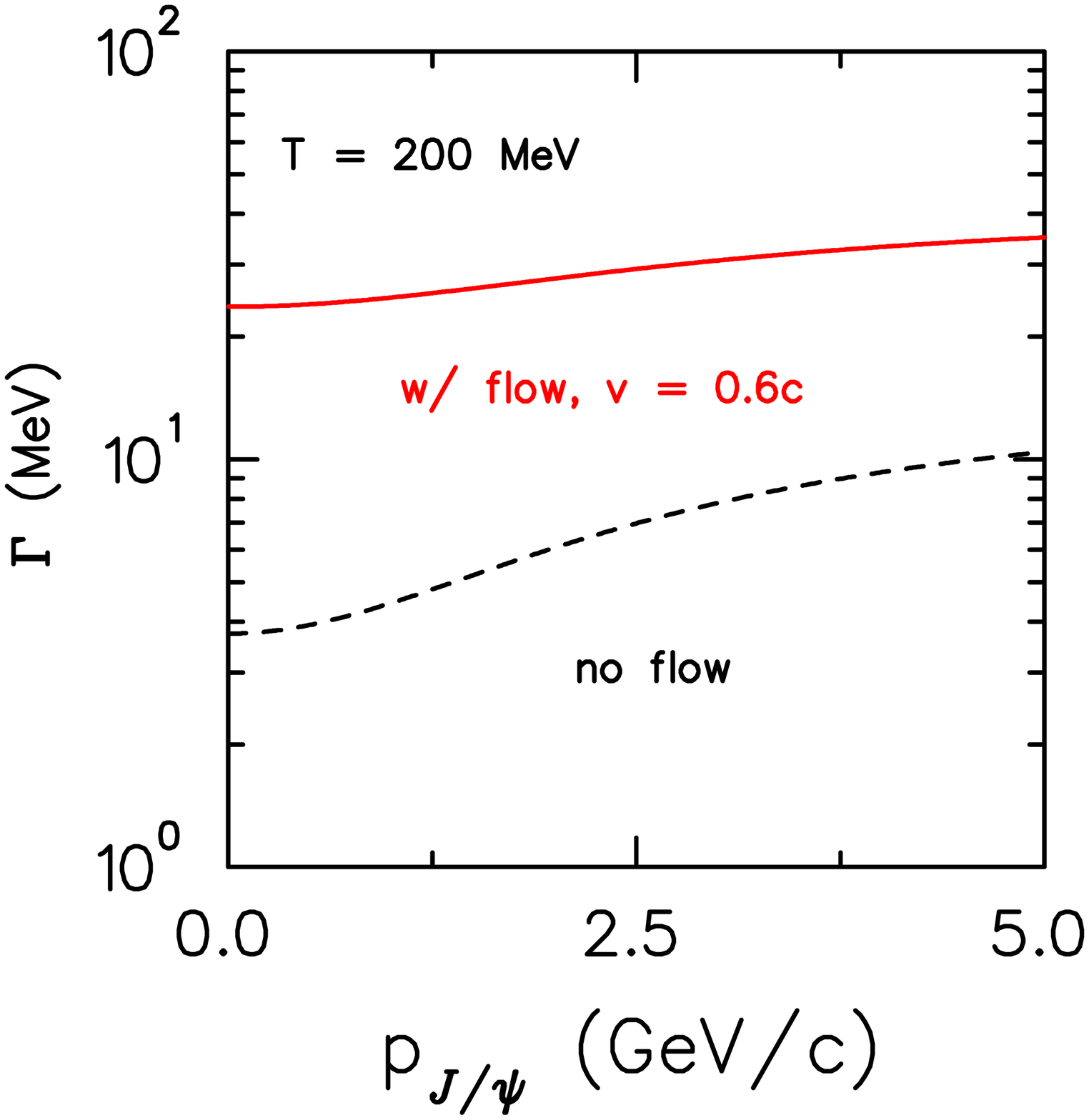}
\caption{\label{fig3}Dissociation rates in the presence of rapid radial flow
$|{\bf v}|$ = 0.6 assuming the $J/\psi$ has fixed three-momentum
randomly distributed.}
\end{figure}

It seems fairly clear by looking at the experimental results from RHIC that
significant flow is present in the reaction zone\cite{dk}.  
Comover suppression
is not expected to be significantly affected if the heavy charm is
comoving.  We suppose here, that it is not.  We look at the possibility
that $J/\psi$ breakup rates could depend on the radial flow velocity.
There is no reason to expect the $J/\psi$s to have thermal 
momentum distributions since the elastic cross sections are
too small to allow complete thermalization\cite{khcg01}.  We therefore
estimate the new dissociation rates assuming rapid radial expansion, 
but with $J/\psi$ given a fixed three-momentum in the rest
frame of the fireball.  We then average over all solid angles
with equal weight. 
The expression we use ``with flow'' is therefore
\begin{equation}
d\/\Gamma_{J/\psi}^{\mbox{wf}} = {d\/\Omega_{{\bf\,p}_{J/\psi}}\over\,4\pi}
\,d\/\Gamma_{J/\psi},
\end{equation}
where $d\Gamma_{J/\psi}$ from Eq.~(\ref{dgamma}) is used but the
equilibrium distribution for particle $b$ is now $f_{eq}(p_{b}\cdot\,U)$,
where $U = \gamma\,(1,{\bf v\/})$ is  the four-velocity.   We use $|{\bf v}|$
= 0.6.   Rates in the presence of flow are reported in Fig.~\ref{fig3}.

\section{Conclusions}\label{concl}
We have estimated charmonium structure using a confining quark potential
calibrated with a long list of hadronic states.  The momentum space
wave functions provide form-factor input to an effective four-flavor
chiral Lagrangian describing charmonium dynamics in a strongly-interacting
many-particle system.  Flow was introduced, albeit in a rather simplistic
way, and was shown to affect the results.

$J/\psi$ breakup cross sections of several tenths to possibly a few 
millibarns were found.  Kinetic theory was used to benchmark the
dissociation rates in a fireball.  At high $J/\psi$ momentum (5 GeV/$c$)
and high system temperature (200 MeV), a dissociation rate of
10 MeV was found.  With flow present in the system, the dissociation
rate increased by a factor of roughly 3.   Future studies will 
include a folding of a more realistic $p_{T}$ distribution for $J/\psi$.
 
\section*{Acknowledgement(s)}
I thank Prof. Charles Gale for useful exchanges regarding this
project.   This work was supported in part by the National Science
Foundation under grant numbers PHY-9814247 and 0098760.

\end{document}